# Boundary Effects on the Determination of Metamaterial Parameters from Normal Incidence Reflection and Transmission Measurements

Sung Kim, *Member, IEEE*, Edward F. Kuester, *Fellow, IEEE*, Christopher L. Holloway, *Fellow, IEEE*, Aaron D. Scher, and James Baker-Jarvis, *Fellow, IEEE*

*Abstract*—**A method is described for the determination of the effective electromagnetic parameters of a metamaterial based only on external measurements or simulations, taking boundary effects at the interfaces between a conventional material and metamaterial into account. Plane-wave reflection and transmission coefficients at the interfaces are regarded as additional unknowns to be determined, rather than explicitly dependent on the material parameters. Our technique is thus analogous to the line-reflect-line (LRL) calibration method in microwave measurements. The refractive index can be determined from S-parameters for two samples of different thickness. The effective wave impedance requires the additional assumption that generalized sheet transition conditions (GSTCs) account for the boundary effects. Expressions for the bulk permittivity and permeability then follow easily. Our method is validated by comparison with the results using the Nicolson-Ross-Weir (NRW) for determining properties of an ordinary material measured in a coaxial line. Utilizing S-parameters obtained from 3-D full wave simulations, we test the method on magnetodielectric metamaterials. We compare the results from our method and the conventional one that does not consider boundary effects. Moreover, it is shown that results from our method are consistent under changes in reference plane location, whereas the results from other methods are not.**

## I. INTRODUCTION

IN recent years, artificial electromagnetic materials have attracted much attention because of their promising applications (e.g., perfect lenses, antennas with improved performance, controllable reflection and transmission devices, electromagnetic absorbers, etc. [1]-[4]). Metamaterials with simultaneously negative permittivity and permeability (variously called double-negative (DNG), negative-refractive-index (NRI), left-handed (LH), backward-wave (BW), Veselago or negative phase velocity (NPV) media) are often needed to achieve these design goals. Such media usually exhibit strong frequency dispersion.

A number of studies on the determination of metamaterial parameters (permittivity, permeability, refractive index and wave impedance) have appeared in recent years [5]-[6]. Accurate extraction of material parameters is very important, because it allows the potential features of metamaterials to be incorporated into a design. A commonly used retrieval method uses S-parameters resulting from a normally incident wave on a metamaterial slab and generates the effective parameters of the metamaterial, assuming that the boundaries of a slab are well defined, and that the Fresnel formulas for reflection and transmission hold at the interface between the air and the metamaterial. However, the tangential components of the *macroscopic* electromagnetic field in a metamaterial are not continuous at the boundaries, although the local fields are. In fact, excess polarization and magnetization due to electric and magnetic multipole moments are induced near the boundary on the scatterers that are constituents of a metamaterial. Therefore, it is difficult to account for the boundaries and the effective length of a metamaterial slab that exhibits the desired bulk properties. Those boundary effects have never been fully investigated in the context of retrieving the material properties of a metamaterial, though some consideration was given them during the early days of the development of artificial dielectrics.

Cohn [7] developed and implemented a method to extract the index of refraction of an artificial delay medium composed of a regular arrangement of conducting obstacles from the measured transmittance of a test slab. The boundary was modeled as a shunt susceptance connected to equivalent transmission lines representing the media, which led to an underdetermined system of equations from which the effective index of refraction was obtained. To circumvent this problem, Cohn assumed the shunt susceptance of the boundary to be one half the susceptance of an isolated plane of scatterers and was forced to make a "judicious estimate" of the wave impedance inside the medium. Brown and Jackson [8] considered an artificial dielectric as a cascaded sequence of T-networks and proposed different models to account for the reactive fields at the interface. The simplest of these models involves simply shifting the position of the effective interface in front of the physical boundary to account for the apparent phase discontinuities in the transmitted and reflected waves. Kharadly [9] used a parallel-plate waveguide to investigate experimentally the properties of artificial dielectrics. In his technique, the effective index of refraction and effective wave impedance were extracted from measured standing wave ratios of test samples terminated by open and short circuits. The author demonstrated that fixing the position of the sharp effective interface relative to the physical interface generated significant experimental error. He went on to treat the position of the effective interface itself as an unknown variable, and solved for it using measured data from two different sample lengths. With this approach, he found considerable improvement in the consistency of the experimental results.

Brown and Seeley [10] went beyond simple transmission line analysis and modeled a metal-strip artificial dielectric as a cascaded series of coupled multiport networks. In this manner, each parallel plane of metal-strips (represented as a unit cell) is connected to another unit cell by multiple transmission lines, each representing a normal mode of the artificial dielectric. In this way, the effects of the evanescent modes excited at the interface can be explicitly taken into account. Using this model, and neglecting all but the least attenuated evanescent mode, the reflection and transmission coefficients at the interface of a semi-infinite artificial delay medium and free space were obtained. The position of the effective interface was then determined by matching the phases of the reflected and transmitted waves calculated for the physical structure to those calculated for an equivalent effective continuous medium.

Outside the aforementioned approaches to the modeling of the boundary for extracting the effective refractive index and wave impedance of an artificial medium, the additional boundary condition (ABC) would need to be mentioned here as the predecessor that considered the transition surface of a crystal structure. The ABC concept was proposed foremost by Pekar [11]. He introduced the additional boundary condition, which assumes that the polarization vanishes at the interface, to intuitively deal with a spatially dispersive material, instead of using the Maxwell's boundary conditions, since it was known that such a classical boundary condition (i.e., the continuity of the tangential components of electromagnetic fields) was insufficient to treat macroscopic electromagnetic fields appearing at the boundary. Later on, Henneberger [12] revisited Pekar's ABC and reformulated the ABC to analytically calculate the reflectivity of the surface of a spatially dispersive medium, accounting for the polariton. Silveirinha et al. took advantage of the ABC for modeling of a wire-composed medium (an array of metallic wires) [13], [14] and mushroom-structured surface [15], which are possible configurations of metamaterials.

For independent approaches of very commonly used extraction method ([5], [6]) based on the Fresnel reflection coefficient, several works were reported. To use the technique studied by Scher et al. [16], assuming that the point-dipole approximation is valid, the electric and magnetic polarizabilities of each sphere are extracted from the S-parameters, and the effective permittivity and permeability are then found by substituting the polarizabilities into the Clausius-Mossotti equations. Simovski et al. [17], [18] reported work related to [16]. In their model, a multipole expansion is initially performed and then a dipole approximation for field interactions of the scatterers is then considered to describe the local permittivity. This method was extended in [19] to find the local permeability. In other work, Simovski et al. [17]-[20] reviewed Drude's quasi-static model to consider the modeling and extraction of the material parameters of a metamaterial, and in [21] and [22] Drude's original idea of a transition layer was extended to apply to the material extraction of a metamaterial for the case where the phase shift across the transition layer between air and metamaterial slab is not negligible. It should be noted that Drude's model is actually a special case of the GSTCs used in this paper (see [23]). The transition layer approach beyond the quasi-static limit ([17]-[19], [21]) can be mathematically equivalent to our GSTC approach. In the theory studied in [17]-[19], [21], the thickness of the layer transition can be a fitting parameter, which is not exactly obtainable. In contrast, the technique presented in this paper is more appropriate for implementation, in that our approach does not require the knowledge of the thickness of the transition layer and the lattice constant $a$, which is more practical. Simovski's transition layer concept aimed to compensate the violation of Maxwell's boundary conditions for macroscopic fields at the interfaces.

In this work, we present a two-sample technique based on the assumption that generalized sheet transition conditions (GSTCs) can be used to describe the jumps in the average (macroscopic) tangential electric and magnetic fields on either side of the interface between a metamaterial and air. Such conditions have previously been used to characterize the field discontinuity across a metafilm: a surface distribution of electrically small scatterers constituting a two-dimensional metamaterial [24]. In the present paper, the GSTCs contain as parameters the excess electric and magnetic surface susceptibilities of the interface, which are in turn dependent on the reflection and transmission coefficients, and the wave impedances of the media. In other words, the jumps of macroscopic fields at the boundary are expressed in the GSTCs by electric and magnetic surface susceptibilities. These susceptibilities help determine effective surface electric and magnetic currents at the interface that approximate the excess polarization and magnetization due to higher-order Bloch modes that are localized near the interfaces. Our work takes into account the boundary effects, which allows us to determine the bulk properties of a metamaterial from measured or simulated scattering data, without the need to have information from the interior of the metamaterial sample.

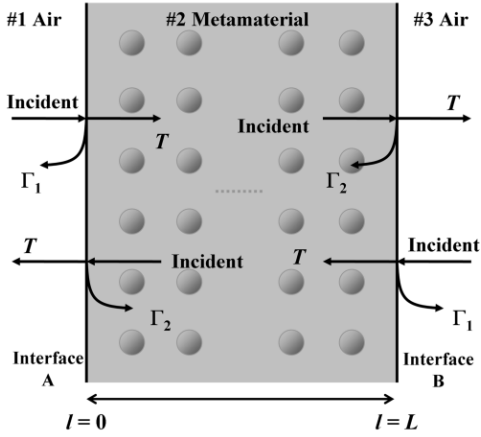

Fig. 1. Schematic illustration of incidence, reflection, and transmission on metamaterial under measurement.

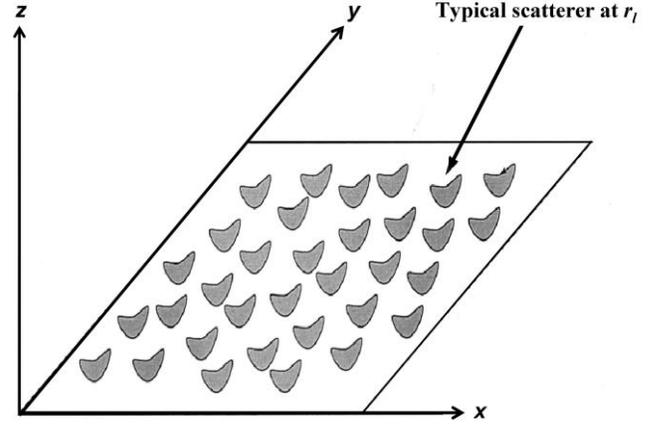

Fig. 2. Two-dimensional metamaterial composed of scatterers distributed in x-y plane.

## II. THEORY

### A. Equations from GSTCs

Consider a slab of metamaterial between two reference planes as shown in Fig. 1. A plane wave normally propagating in air is incident from either side. Several assumptions are made for this configuration. The reflection coefficients $\Gamma_1$, $\Gamma_2$, and the transmission coefficient $T$ are defined as S-parameters with respect to reference planes at Interface A ($l = 0$) and Interface B ($l = L$). These coefficients are not assumed to be related to the bulk material parameters by the Fresnel formulas, but are unknowns to be determined separately. Interfaces A and B are assumed to be reciprocal but not symmetric: the air-to-slab and slab-to-air reflection coefficients ($\Gamma_1$ and $\Gamma_2$) are different in general, whereas the transmission coefficients in both directions are the same.

Now, we will make use of GSTCs of second order, similar to those in [25], as a means of obtaining an additional relationship between $\Gamma_1$, $\Gamma_2$, and $T$ without using the Fresnel formulas. In previous work, Kuester *et al.* [24] studied these averaged transition conditions for the average or macroscopic electromagnetic fields at a metafilm. These boundary conditions contain electric and magnetic surface susceptibilities of the metafilm as parameters. There is evidence to suggest that GSTCs of this type also govern the macroscopic fields at the interface between a metamaterial and an ordinary medium [26], so long as higher-order Bloch modes in the metamaterial are all evanescent and thus localized near the interfaces. These GSTCs help determine reflection and transmission at the interface, accounting for boundary effects due to excess surface polarization and magnetization induced at the scatterers. In Fig. 2, the GSTC equations are represented by

$$\vec{a}_z \times \vec{H} \mid_{z=0-}^{0+} = j\omega\varepsilon\bar{\bar{\chi}}_{ES} \cdot \vec{E}_{t,av} \mid_{z=0} - \vec{a}_z \times \nabla_t \left[ \bar{\bar{\chi}}_{MS}^{zz} H_{z,av} \right]_{z=0} \quad (1)$$

$$\vec{E} \mid_{z=0-}^{0+} \times \vec{a}_z = j\omega\mu\bar{\bar{\chi}}_{MS} \cdot \vec{H}_{t,av} \mid_{z=0} - \nabla_t \left[ \chi_{ES}^{zz} E_{z,av} \right]_{z=0} \times \vec{a}_z \quad (2)$$

where $\vec{E}_{av}$ and $\vec{H}_{av}$ are the average macroscopic electromagnetic fields on either side of the interface, and $\bar{\bar{\chi}}_{ES}$ and $\bar{\bar{\chi}}_{MS}$ are effective dyadic electric and magnetic surface susceptibilities. The GSTC presented in [24] was not the most general one possible, in that it was assumed that two media on both sides of the interface are identical. Hence, (1) and (2) may need to include respectively additional terms for the jumps of electric and magnetic fields at the boundary to deal with a non-symmetrical interface. However, the numerical calculation we have performed using 3-D full wave simulation as a preparatory exercise has showed that the GSTCs in [24] provide reasonable values for $\chi_{ES}$ and $\chi_{MS}$ in comparison with non-symmetric GSTCs. This verification allows us to make immediate use of (1) and (2). Since there is no way to determine the degree of asymmetry present in the GSTCs from purely external information, this simplification is greatly beneficial to our technique.

Let Interface A be located in the $z = 0$ in Fig. 2. Assume a plane wave normally incident on Interface A propagates in the direction of $z$ axis from the air to the metamaterial. The electric and magnetic fields in $z < 0$ are expressed such as

$$\vec{E} = \vec{a}_y \left( e^{-jk_0 z} + \Gamma_1 e^{jk_0 z} \right) \quad (3)$$

$$\vec{H} = \vec{a}_x \frac{1}{\varsigma_0} \left( -e^{-jk_0 z} + \Gamma_1 e^{jk_0 z} \right) \quad (4)$$

and in $z > 0$,

$$\vec{E} = \vec{a}_y T_1 e^{-jk_{eff} z} \quad (5)$$

$$\vec{H} = -\vec{a}_x \frac{T_1}{\varsigma_{eff}} e^{-jk_{eff} z} \quad (6)$$

where $k_0$ and $k_{eff}$ are the wave numbers of the air and metamaterial and $\varsigma_0$ and $\varsigma_{eff}$ are the (absolute) wave impedance of the air and metamaterial. Note that in (5) and (6), $T_1$ is a voltage-ratio transmission coefficient rather than an S-parameter and so that $T_1 = T\sqrt{\varsigma_{eff}/\varsigma_0}$ where $T$ is an S-parameter. Substituting (3)-(6) into (2) and (1) gives the effective electric and magnetic surface susceptibilities at Interface A in Fig 1 respectively:

$$\chi_{ES,A}^{yy} = \frac{\dfrac{1}{\varsigma_0} - \dfrac{\Gamma_1}{\varsigma_0} - \dfrac{T}{\sqrt{\varsigma_{eff}\,\varsigma_0}}}{\dfrac{1}{2}\,j\omega\varepsilon_0\left(1 + \Gamma_1 + T\sqrt{\dfrac{\varsigma_{eff}}{\varsigma_0}}\right)} \quad (7)$$

$$\chi_{MS,A}^{xx} = \frac{T\sqrt{\dfrac{\varsigma_{eff}}{\varsigma_0}} - 1 - \Gamma_1}{\dfrac{1}{2}\,j\omega\mu_0\left(\dfrac{1}{\varsigma_0} - \dfrac{\Gamma_1}{\varsigma_0} + \dfrac{T}{\sqrt{\varsigma_{eff}\,\varsigma_0}}\right)} \quad (8)$$

Furthermore, if we consider the case when a plane wave propagates from the metamaterial to the air, we can obtain the effective electric and magnetic surface susceptibilities at Interface B as follows,

$$\chi_{ES,B}^{yy} = \frac{\dfrac{1}{\varsigma_{eff}} - \dfrac{\Gamma_2}{\varsigma_{eff}} - \dfrac{T}{\sqrt{\varsigma_{eff}\,\varsigma_0}}}{\dfrac{1}{2}\,j\omega\varepsilon_0\left(1 + \Gamma_2 + T\sqrt{\dfrac{\varsigma_0}{\varsigma_{eff}}}\right)} \quad (9)$$

$$\chi_{MS,B}^{xx} = \frac{T\sqrt{\dfrac{\varsigma_0}{\varsigma_{eff}}} - 1 - \Gamma_2}{\dfrac{1}{2}\,j\omega\mu_0\left(\dfrac{1}{\varsigma_{eff}} - \dfrac{\Gamma_2}{\varsigma_{eff}} + \dfrac{T}{\sqrt{\varsigma_{eff}\,\varsigma_0}}\right)} \quad (10)$$

Notice that the metamaterial slab has been modeled here as a continuous medium with (7)-(10) describing the interface (boundary) effects separately from the bulk properties of refractive index and wave impedance. No assumption has been made about an averaging method (surface, volume, etc.) either inside the metamaterial layer or on its boundary, since such information would not easily be available from measured data.

In this paper, the metamaterial structures of practical interest possess symmetry along the propagation direction. Moreover, the effective surface susceptibilities at the interfaces are assumed to be same as for a semi-infinite medium. Therefore, the layer thicknesses must be assumed large enough that near-field interaction is negligible. In such a case, we let both $\chi_{ES,A}^{yy} = \chi_{ES,B}^{yy}$ in (7) and (9) and $\chi_{MS,A}^{xx} = \chi_{MS,B}^{xx}$ in (8) and (10). This leads to identical expressions for the effective wave impedance of the metamaterial slab:

$$\varsigma_{eff} = \sqrt{\frac{\mu_{eff}}{\varepsilon_{eff}}} = \frac{-\Gamma_1 + \Gamma_2 + \Gamma_1\Gamma_2 - T^2 - 1}{\Gamma_1 - \Gamma_2 + \Gamma_1\Gamma_2 - T^2 - 1}\,\varsigma_0$$
$$= \frac{A + t\left(1 + \Gamma_1\right)}{B + t\left(1 - \Gamma_1\right)}\,\varsigma_0 \quad (11)$$

where

$$A = t\Gamma_2 + \Gamma_1\left(t\Gamma_2\right) - tT^2$$
$$B = -t\Gamma_2 + \Gamma_1\left(t\Gamma_2\right) - tT^2, \quad (12)$$

which can be found from the S-parameters if $\Gamma_1$, $t\Gamma_2$, and $tT^2$ are expressed in terms of S-parameters. The quantity $t$ in (11) and (12) is defined in (15) in the next section. It is important to note that if $A$ and $B$ are assumed to be zero in (11), then the expression for the wave impedance reduces to that of the Fresnel reflection coefficient, $\varsigma_{eff} = \varsigma_0\left(1 + \Gamma_1\right)/\left(1 - \Gamma_1\right)$.

### B. Equations from Measurement

Together with the assumption made in the previous section that Interfaces A and B are reciprocal but not symmetric, if it is assumed that the material structure possesses symmetry with respect to the wave propagation direction, and that the measured or simulated S-parameters obey $S_{11} = S_{22}$ and $S_{21} = S_{12}$, then by considering multiple reflections of an incident wave bouncing between Interface A and B or by using a signal flow graph for Fig. 1, as explained in [27]-[29], the following equations are obtained:

$$S_{21} = \frac{tT^2}{1 - \left(t\Gamma_2\right)^2} \quad (13)$$

$$S_{11} = \Gamma_1 + tS_{21}\Gamma_2 \quad (14)$$

where

$$t = \exp\left[-jk_0 n_{eff} L\right] = \exp\left(\pm j\omega\sqrt{\varepsilon_{eff}\,\mu_{eff}}\,L\right), \quad (15)$$

while $S_{21}$ and $S_{11}$ are the measured or simulated S-parameters, $n_{eff}$ is the effective refractive index, and $\varepsilon_{eff}$ and $\mu_{eff}$ are the effective permittivity and permeability. If the medium is passive, the correct sign in (15) must be such that $\text{Im}\left(n_{eff}\right) \leq 0$. Finally, $L$ is the effective length of the slab. Determination of the absolute physical slab length may also be of interest for metamaterial studies. Nonetheless, we provisionally define $L = Na$, letting $N$ be the number of layers (unit cells) and $a$ the lattice constant (unit cell size) in the direction perpendicular to the slab. The effects due to varying the reference plane positions on the resultant metamaterial properties will be examined in Section III *B* and

*C.*

Inserting the S-parameters extracted from the measurement or simulation for samples of two different lengths $L = L1$ and $L = L2$ into (13)-(15) provides 6 equations with 6 unknowns ( $\Gamma_1$, $\Gamma_2$, $T$, $t^{L1}$, $t^{L2}$, and $n_{eff}$ ). From those equations, the coefficients $\Gamma_1$, $t^{L1}\Gamma_2$, and $t^{L1}T^2$ can be solved for analytically as functions of the S-parameters as follows:

$$\Gamma_1 = \frac{-X \pm \sqrt{X^2 - 4\left(S_{11}^{L1} - S_{11}^{L2}\right)Y}}{2\left(S_{11}^{L1} - S_{11}^{L2}\right)} \quad (16)$$

where

$$X = -\left(S_{11}^{L1}\right)^2 + \left(S_{11}^{L2}\right)^2 + \left(S_{21}^{L1}\right)^2 - \left(S_{21}^{L2}\right)^2$$
$$Y = \left(S_{11}^{L1}\right)^2 S_{11}^{L2} - S_{11}^{L1}\left(S_{11}^{L2}\right)^2 + S_{11}^{L1}\left(S_{21}^{L2}\right)^2 - S_{11}^{L2}\left(S_{21}^{L1}\right)^2 \quad (17)$$

and

$$t^{L1}\Gamma_2 = \frac{S_{11}^{L1} - \Gamma_1}{S_{21}^{L1}} \quad (18)$$

$$t^{L1}T^2 = S_{21}^{L1}\left\{1 - \left(\frac{S_{11}^{L1} - \Gamma_1}{S_{21}^{L1}}\right)^2\right\} = S_{21}^{L1}\left\{1 - \left(t^{L1}\Gamma_2\right)^2\right\}. \quad (19)$$

The superscripts $L1$ and $L2$ in (16)-(19) denote parameters obtained from measurements or simulations with the corresponding sample lengths. Interchanging $S_{21}^{L1}$, $S_{11}^{L1}$, and $t^{L1}$ with $S_{21}^{L2}$, $S_{11}^{L2}$, and $t^{L2}$ yields the same values of $\Gamma_2$ and $T^2$ in (18) and (19). As is seen in (19), it is possible to find $T^2$ but not $T$ by this method. It is worth noting here that (16)-(19) are similar to the equations of the line-reflect-line (LRL) calibration method [30]-[37]. This calibration method is used to determine repeatable measurement errors on both sides of a two-port device under test. The formulation in the calibration method allows one to extract a complex propagation constant from the measurement of a transmission line or waveguide of two different lengths. In this context, Interfaces A and B in Fig. 1 correspond to the error boxes in the LRL calibration method.

A metamaterial without active constituents is a passive medium. The reflection coefficient $\Gamma_1$ must therefore have a magnitude less than or equal to unity in such cases, with the air medium considered to be lossless. This condition can often be used to resolve the sign ambiguity of $\Gamma_1$ in (16). However, the magnitude of the reflection coefficient $\Gamma_2$ may sometimes exceed unity, if the metamaterial under investigation is lossy [38].

Once $\Gamma_1$, $t^{L1}\Gamma_2$, and $t^{L1}T^2$ are obtained, the effective refractive index of the metamaterial is obtained from (14) and (15) as

$$n_{eff} = \frac{j}{k_0\left(L2 - L1\right)}\left[\ln\left\{\frac{S_{21}^{L1}\left(S_{11}^{L2} - \Gamma_1\right)}{S_{21}^{L2}\left(S_{11}^{L1} - \Gamma_1\right)}\right\}\right]$$
$$= \frac{j}{k_0\left(L2 - L1\right)}\left[Ln\left\{\frac{S_{21}^{L1}\left(S_{11}^{L2} - \Gamma_1\right)}{S_{21}^{L2}\left(S_{11}^{L1} - \Gamma_1\right)}\right\} + j2\pi m\right] \quad (20)$$

where $m$ is an integer. Note that we must have $\text{Im}\left(n_{eff}\right) \le 0$ for a passive medium. This physical requirement could be imposed in such a way that a correct sign is found for the reflection coefficient $\Gamma_1$ in (16). The real part of the refractive index in (20) involves choosing a branch for the logarithm, indicated by the integer $m$. If the group delay of $S_{21}$ were smooth over the frequency range of interest, it could be utilized to determine $m$ [39]-[43]. The phase unwrapping method [44] is another way of resolving this ambiguity. It is our experience, however, that most metamaterials are very dispersive, and the group delay changes rapidly around resonance frequencies, making the choice of $m$ a more difficult problem.

Another problem we have encountered is that a metamaterial is often very lossy in the frequency bands of interest. The coefficients calculated from (16) and (18) can be very sensitive to the measured or simulated S-parameters in this case, when almost no transmission through the slab occurs. To help address this problem, we assume that $S_{21}^{L1}$ and $S_{21}^{L2}$ are very small, and introduce the differences and averages of S-parameters as new parameters:

$$\Delta S_{11} = S_{11}^{L1} - S_{11}^{L2}$$
$$S_{11}^{av} = \frac{S_{11}^{L1} + S_{11}^{L2}}{2}$$
$$\Delta S_{21} = S_{21}^{L1} - S_{21}^{L2}$$
$$S_{21}^{av} = \frac{S_{21}^{L1} + S_{21}^{L2}}{2} \quad (21)$$

Substitution of the quantities in (21) into (16)-(18) result in the following modified expressions for the reflection and transmission coefficients:

$$\Gamma_1 = S_{11}^{av} - S_{21}^{av}\frac{\Delta S_{21}}{\Delta S_{11}}$$
$$\pm \frac{\Delta S_{11}}{2}\sqrt{\left[1 - \left(\frac{\Delta S_{21}}{\Delta S_{11}}\right)^2\right]\left[1 - 4\left(\frac{S_{21}^{av}}{\Delta S_{11}}\right)^2\right]} \quad (22)$$

$$t^{L1}\Gamma_2 = \frac{\dfrac{S_{21}^{av}}{\Delta S_{11}} + \dfrac{\Delta S_{21}}{2\Delta S_{11}}}{\dfrac{1}{2} + \dfrac{S_{21}^{av}\Delta S_{21}}{\left(\Delta S_{11}\right)^2} \pm \dfrac{1}{2}\sqrt{\left[1 - \left(\dfrac{\Delta S_{21}}{\Delta S_{11}}\right)^2\right]\left[1 - 4\left(\dfrac{S_{21}^{av}}{\Delta S_{11}}\right)^2\right]}} \quad (23)$$

Equation (19) remains unchanged, but $t^{L1}\Gamma_2$ computed from

(23) should be used in (19) in place of (18). Equations (22) and (23) are the functions of the ratios $\Delta S_{21}/\Delta S_{11}$ and $S_{21}^{av}/\Delta S_{11}$. If terms of second order in these ratios are much smaller than the other terms, the square roots in (22) and (23) can simply be approximated by unity. We have observed that metamaterials may have very small transmission properties ($S_{21}$) at their resonance frequencies which are almost stopbands. We have also seen that in many cases, the use of (19) and (22)-(23) can help to remove some undesired noise of the results of the retrieved material properties, which may be artificially caused by very small transmission. Therefore, we will use (19) and (21)-(23) rather than (16)-(19) in the following sections for obtaining the results from our method presented in this paper. Furthermore, $t^{L1}$ can be expressed in terms of the ratios as follows:

$$t^{L1} = \exp\left[\frac{L1}{L2-L1}\left\{\ln\frac{S_{21}^{L1}}{S_{21}^{L2}}\right.\right.$$
$$\left.\left.+\ln\frac{-1+2\frac{S_{21}^{av}}{\Delta S_{11}}\frac{\Delta S_{21}}{\Delta S_{11}}\mp\sqrt{\left[1-\left(\frac{\Delta S_{21}}{\Delta S_{11}}\right)^2\right]\left[1-4\left(\frac{S_{21}^{av}}{\Delta S_{11}}\right)^2\right]}}{-1+2\frac{S_{21}^{av}}{\Delta S_{11}}\frac{\Delta S_{21}}{\Delta S_{11}}\mp\sqrt{\left[1-\left(\frac{\Delta S_{21}}{\Delta S_{11}}\right)^2\right]\left[1-4\left(\frac{S_{21}^{av}}{\Delta S_{11}}\right)^2\right]}}\right\}\right]$$

(24)

Now, we are ready to find the effective wave impedance $\varsigma_{eff}$ from the S-parameters by using (11), (12), (19), (22), and (23). It is known that the bulk relative permittivity and permeability are given by $n_{eff}/(\varsigma_{eff}/\varsigma_0)$ and $n_{eff}(\varsigma_{eff}/\varsigma_0)$, respectively. Therefore, from (11) and (20) we have

$$\varepsilon_{r,eff} = \frac{j}{k_0(L2-L1)}\cdot\frac{B^{L1}+t^{L1}(\Gamma_1-1)}{A^{L1}-t^{L1}(\Gamma_1+1)}$$
$$\cdot\left[Ln\left\{\frac{S_{21}^{L1}(S_{11}^{L2}-\Gamma_1)}{S_{21}^{L2}(S_{11}^{L1}-\Gamma_1)}\right\}+j2\pi m\right]$$

(25)

and

$$\mu_{r,eff} = \frac{j}{k_0(L2-L1)}\cdot\frac{A^{L1}-t^{L1}(\Gamma_1+1)}{B^{L1}+t^{L1}(\Gamma_1-1)}$$
$$\cdot\left[Ln\left\{\frac{S_{21}^{L1}(S_{11}^{L2}-\Gamma_1)}{S_{21}^{L2}(S_{11}^{L1}-\Gamma_1)}\right\}+j2\pi m\right],$$

(26)

which are determined (except for the integer $m$) from experimentally or numerically extracted data.

The effective material parameters for a metamaterial are given completely by (11), (20), (25), and (26). The material property determination method presented here is based on measured or simulated data of samples of two different effective lengths. Therefore, it is important that one should take enough layers in each sample in order that the metamaterial can exhibit well-converged bulk material properties. Furthermore, when the metamaterial is subjected to an incident electromagnetic wave with a wavelength that is sufficiently larger than the lattice constant (e. g., $|k_0 n_{eff} a| < 1$ at the frequencies away from a resonance of the metamaterial where $n_{eff} \approx 0$), the metamaterial can then be regarded as a homogeneous effective medium. If this criterion is violated, the correctness of the effective material parameters generated by (11), (20), (25), and (26) cannot be guaranteed.

It should be noted here that our method in the present form assumes that surfaces of the slab are identical. This means that the particles are assumed to be both individually symmetrical and symmetrically arranged in space in the direction normal to the surface (i.e., about the transverse plane). Asymmetrical arrangements of particles, such as the "classic" cascade connection of rods and split-ring resonators along the direction of propagation are not covered by our method in its present form.

### III. RESULTS AND DISCUSSION

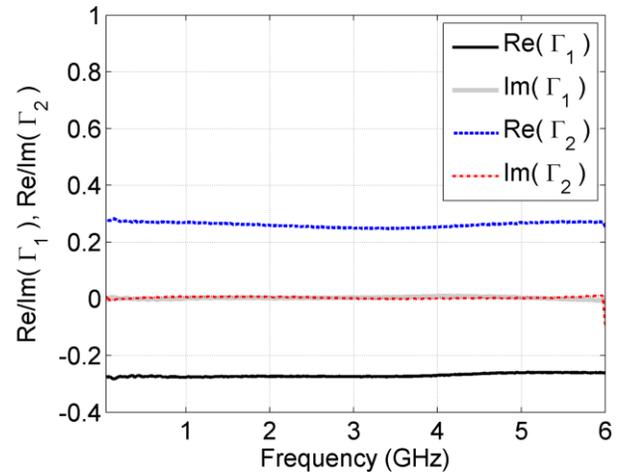

Fig. 3. Real (solid black) and imaginary (solid gray) parts of reflection coefficients $\Gamma_1$ and real (dashed blue) and imaginary (dashed red) parts of $\Gamma_2$ of the nylon sample as a function of frequency.

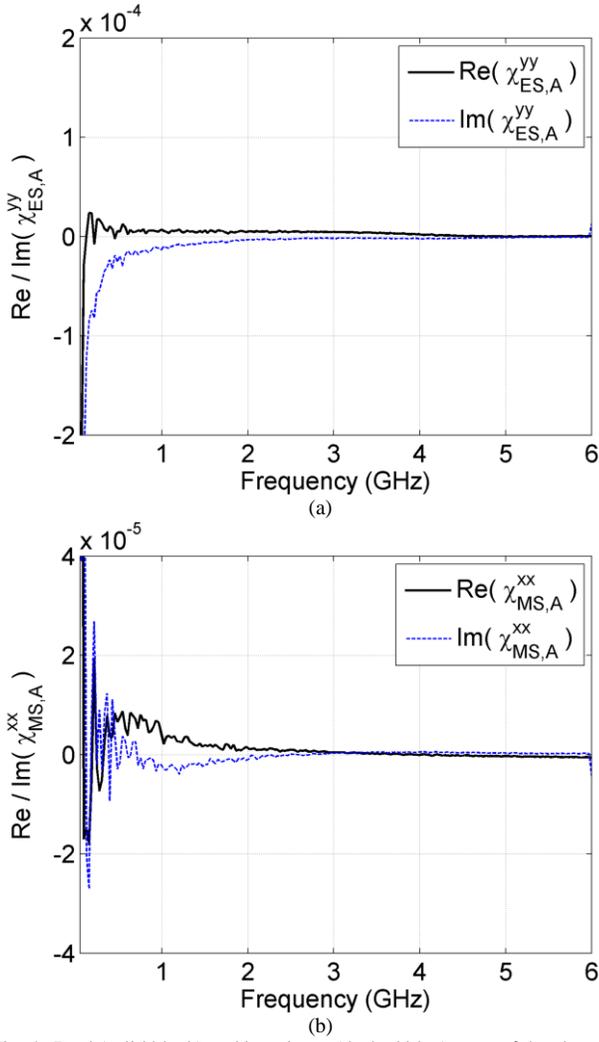

Fig. 4. Real (solid black) and imaginary (dashed blue) parts of the electric and magnetic surface susceptibilities at Interface A: (a) electric surface susceptibilities. (b) magnetic susceptibilities.

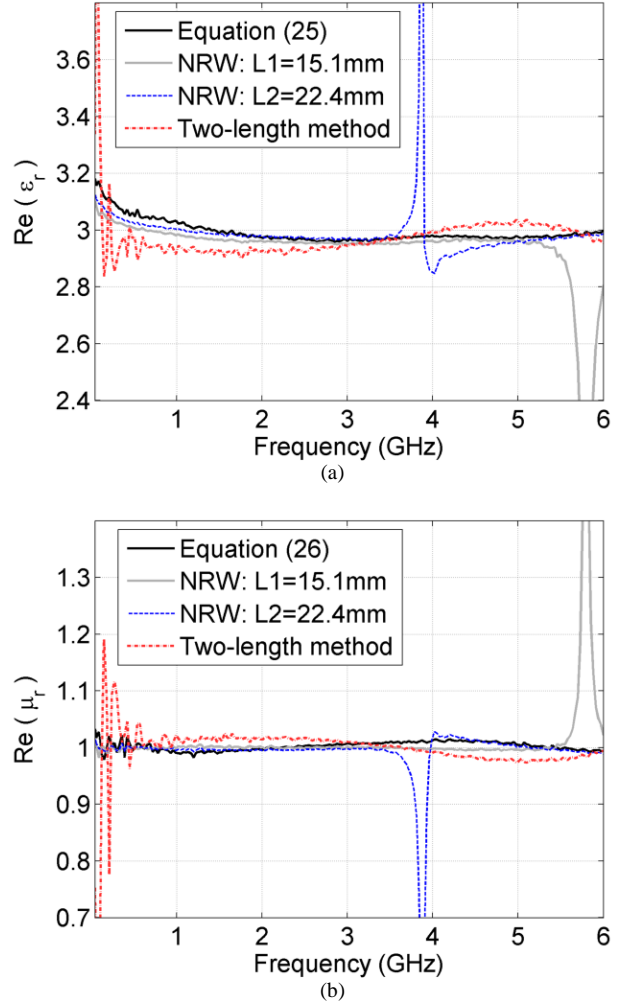

Fig. 5. Comparisons of the relative permittivity and permeability of the nylon sample by use of our (solid black), the NRW (solid gray and dashed blue) equations, and two-length method (dashed red): (a) real part of relative permittivity. (b) real part of relative permeability.

### A. Experimental Results from the Conventional Material

We first validate our method using S-parameters measured from samples of a conventional material. In this case, the boundary effects at the interface between the air and material are expected to be negligible, and the Fresnel relations at the interface to be correct. We used nylon ($\varepsilon_r \approx 3.0$, $\mu_r \approx 1.0$) as the test material. Two different sample lengths, $L1 = 15.1$ mm and $L2 = 22.4$ mm, were prepared. Each sample was placed in a 50 mm long coaxial sample holder that supports transverse electromagnetic (TEM) propagation. This sample holder was introduced between the reference planes at the ends of the cables of vector network analyzer (VNA). Calibration was performed to determine the reference planes. The measurement frequency was varied from 50 MHz to 6 GHz. The S-parameters utilized in our equations were phase-shifted compared with the measured S-parameters to account for the difference in length between the samples and the holder.

First, we obtained the reflection coefficients $\Gamma_1$ from (22)

and $\Gamma_2$ from (23), having found $t^{L1}$ from (24). Fig. 3 shows the real and imaginary parts of $\Gamma_1$ and $\Gamma_2$. It is seen that $\Gamma_1 = -0.27$ and $\Gamma_2 = 0.27$, thus with small discrepancy $\Gamma_1 = -\Gamma_2$ as expected. This indicates that although no assumption about $\Gamma_1$ and $\Gamma_2$ has been made *a priori*, the reflection coefficients in Fig. 3 agree with those from widely used equations, i.e., $\Gamma_1 = -\Gamma_2 = \left( \sqrt{\mu_r / \varepsilon_r} - 1 \right) / \left( \sqrt{\mu_r / \varepsilon_r} + 1 \right)$, where $\varepsilon_r$ and $\mu_r$ are the relative permittivity and permeability of the material under test.

The values for $\chi_{ES,A}$ and $\chi_{ES,B}$ were found to be graphically indistinguishable, and likewise for $\chi_{MS,A}$ and $\chi_{MS,B}$, as expected. Figs. 4(a) and (b) show the electric and magnetic surface susceptibilities induced at Interface A from (7) and (8). It is observed that the values of the electric and magnetic surface susceptibilities are small: $\chi_{ES}^{yy}$ is on the order of $10^{-6}$, and $\chi_{MS}^{xx}$ is on the order of $10^{-7}$. It is thus confirmed that the boundary effects are negligible for a conventional material.

Figs. 5(a) and (b) show comparisons among the real parts of the relative permittivity and permeability from (25) and (26), the Nicolson-Ross-Weir (NRW) procedure [27], [28] and the two-length method presented in [45] (not to be confused with the present method, which also uses two different sample lengths). The NRW method for determining the permittivity and permeability of a material from the reflection and transmission measurements of a single sample is well-known. The NRW equations are derived by using the relation $T = 1 + \Gamma$ where $T$ and $\Gamma$ are the transmission and reflection coefficients of a wave incident from the air onto the sample. NRW results for both sample lengths $L1$ and $L2$ are plotted in Fig. 5. It is known that if a low-loss material is measured, the NRW algorithm can yield spurious peaks at frequencies where the sample length is an integer multiple of one-half wavelength of the material. The reflection/transmission principles on which the NRW method is based also motivated the two-length method [45] to suppress the unwanted peaks seen in the NRW, if the S-parameters on the samples of two different lengths are available. In this method, only the values of $S_{21}^{L1}$ and $S_{21}^{L2}$ are used, and an iterative solution of the relevant equations at each frequency is used. The two-length method introduces more uncertainty than does the NRW algorithm. Other types of two-length approaches are found in the literature [46].

In Fig. 5(a), the real part of permittivity obtained from our method and the NRW method increases somewhat at lower frequencies, but approaches a constant limit $\varepsilon_{r,eff} = 2.96$ as the frequency is increased. In contrast, the two-length method produces large discrepancies with the other approaches at the low frequency region, since the difference between the phases of $S_{21}^{L1}$ and $S_{21}^{L2}$ is not large enough and increases the uncertainty in this frequency region. At frequencies up to about 1.5 GHz, some discrepancy between the results from our method and the NRW method are observed as well. This can be explained by the fact that an air gap between the sample and the coaxial fixture exists, causing experimental error. The real part of permeability, as is seen in Fig. 5(b), shows good agreement with the expected result $\mu_{r,eff} = 1.00$.

The permittivity and permeability computed from the NRW equations show divergences at 5.79 GHz for $L1 = 15.1$ mm and at 3.87 GHz for $L2 = 22.4$ mm. This is because $L1$ and $L2$ are an integer multiple of one-half wavelength in the material at those frequencies. With our technique, by keeping the difference between the two sample lengths $L2 - L1$ sufficiently small we can eliminate the divergences in the frequency range of interest. The results from our equations will diverge at frequencies where $S_{11}^{L1}$ and $S_{11}^{L2}$ are simultaneously small. Numerical calculations indicate that the peaks in the extracted permittivity and permeability of the nylon will occur at 11.92, 23.84, 35.76 GHz, etc.

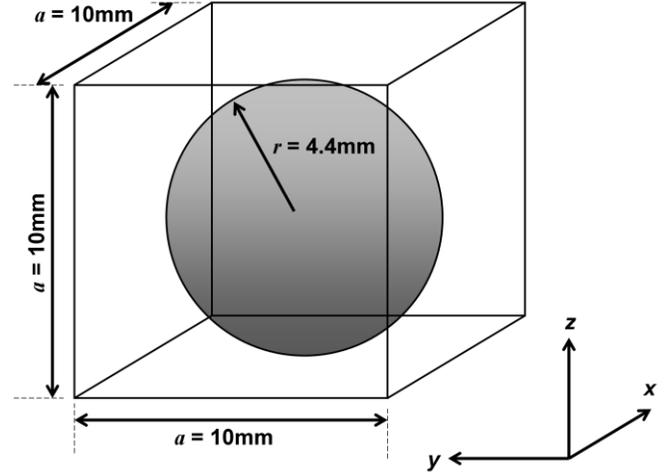

Fig. 6. Unit cell for a cubic array of magnetodielectric spheres.

## B. Case 1: Simulation Results from a Cubic Array of Magnetodielectric Spheres with $\varepsilon_r = \mu_r$

Next, we investigated the effective properties using numerically simulated S-parameters from metamaterial slabs. A cubic array of magnetodielectric spheres were studied in our 3-D full wave simulations. This structure possesses the requisite symmetry (recall that we assumed that $S_{11} = S_{22}$ and $S_{21} = S_{12}$ as well as isotropy of the metamaterial in deriving the equations in this paper). This kind of structure has been studied and shown to exhibit negative material properties due to the resonant spherical inclusions [2], [43], [47].

The simulations were performed with the finite-element-based software, Ansoft's HFSS (High Frequency Structure Simulator). Fig. 6 shows the structure of the cubic unit cell used in the simulations. The values of $r = 4.4$ mm for the radius of the spheres and $a = 10$ mm for the lattice constant were chosen to have a resonance at our desired frequency.

Perfect electric conductor (PEC) boundary conditions were assigned for the top and bottom walls (x-y planes) of the unit cell for a vertically-polarized electric field. Perfect magnetic conductor (PMC) boundary conditions were employed for the side walls (z-x planes) for a horizontally-polarized magnetic field. This corresponds to normal incidence (x) of a TEM plane wave. In the x direction, the unit cell was repeated in such a way that a suitable number of layers in the sample were present.

We examined the case when the permittivity and permeability of the inclusions was the same: $\varepsilon_r = \mu_r = 50 - j0.05$. Theoretical results predict that boundary effects in this case should not be very significant [48]. These parameters produce a bulk effective wave impedance identical to that of air. Two samples with differing numbers of layers (unit cells), $N1 = 9$ and $N2 = 10$, were chosen for this configuration.

Figs. 7(a) and (b) show the magnitudes of the reflection coefficients $\Gamma_1$ and $\Gamma_2$ obtained from (22) and (23). For this metamaterial, $\Gamma_1$ and $\Gamma_2$ differ from each other in magnitude, unlike for the conventional material discussed in Section III *A*. This demonstrates that the assumption, $\Gamma_1 = -\Gamma_2$, as is often made in other extraction techniques based on the Fresnel formulas, is not necessarily true.

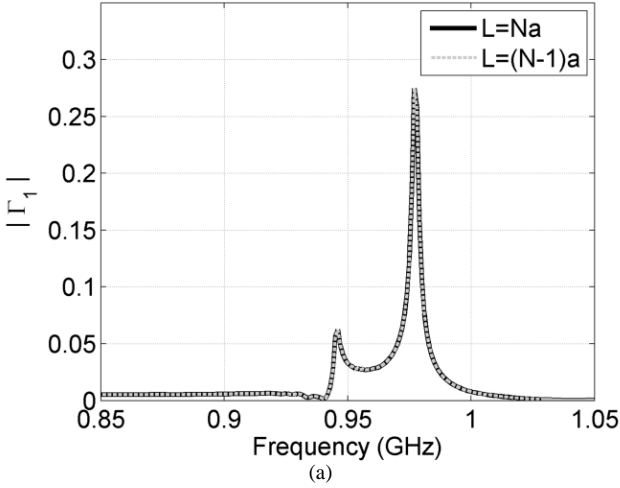

(a)

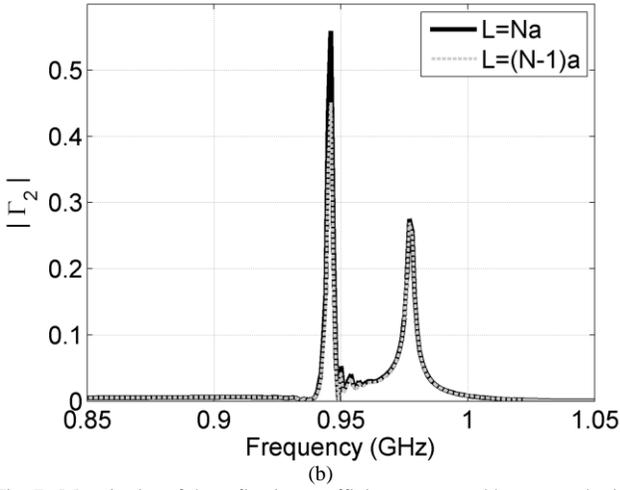

(b)

Fig. 7. Magnitudes of the reflection coefficients extracted by our method for $\varepsilon_r = \mu_r = 50 - j0.05$ with $L = Na$ (solid black) and $L = (N-1)a$ (dashed gray): (a) magnitude of $\Gamma_1$. (b) magnitude of $\Gamma_2$.

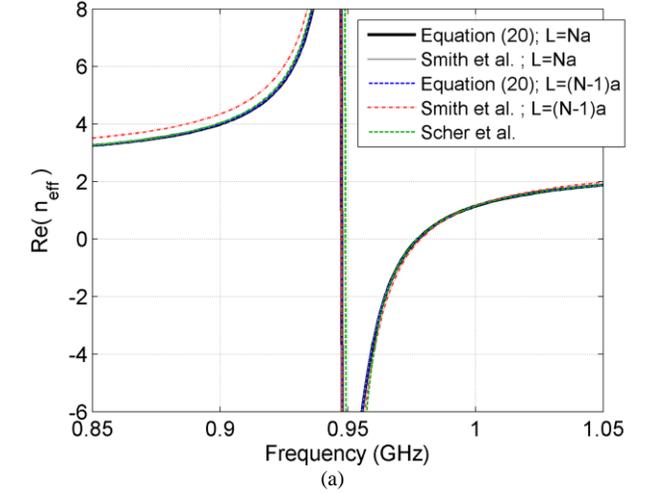

(a)

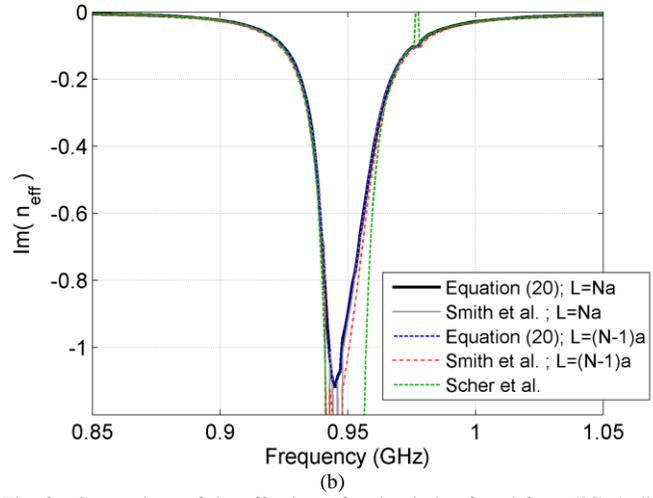

(b)

Fig. 8. Comparison of the effective refractive index found from (20) (solid black and dashed blue), from the equations of Smith *et al.* (solid gray and dashed red) and from the method of Scher *et al.* (dashed green) for $\varepsilon_r = \mu_r = 50 - j0.05$ with $L = Na$ and $L = (N-1)a$: (a) real part of $n_{eff}$, (b) imaginary part of $n_{eff}$.

The cubic array of magnetodielectric spheres was designed to have a resonance around 0.949 GHz. The peaks in $|\Gamma_1|$ and $|\Gamma_2|$ shown in Figs. 7(a) and (b) occur at 0.946 and 0.977 GHz, and demonstrate that the slab-to-air propagation has a strong reflectivity at the resonance frequency compared with the air-to-slab propagation. It is also to be noted that by use of our equations, $|\Gamma_1|$ is consistent and independent of the choice of reference planes, i.e., $L = Na$ or $L = (N-1)a$ (the results are graphically indistinguishable), since our method of finding the reflection and transmission coefficients uses two different effective sample lengths, so that the reference plane shift is

cancelled, if the same shift is used for each length.

To compare these results with those from other methods discussed in the literature, we first use the equations from Smith *et al.* [5]. Their effective property retrieval method was based on a transfer matrix containing information on effective refractive index, wave impedance, and material length for a homogeneous material, similar in spirit to the NRW method and requiring only the S-parameters of a single effective slab length. In implementing their method, we used data for a slab ten layers thick. Also, results from Scher's method [16] for the metamaterial with the configuration described here are available.

The effective refractive index resulting from (20) is presented in Fig. 8. In Fig. 8(a), the real part of the effective refractive index switches its sign at 0.949 GHz and becomes negative. In Fig. 8(b), the imaginary part shows a physical value for a passive medium over the entire frequency range: Im $[n_{eff}] \leq 0$ . If $L = Na$ is chosen, the refractive index

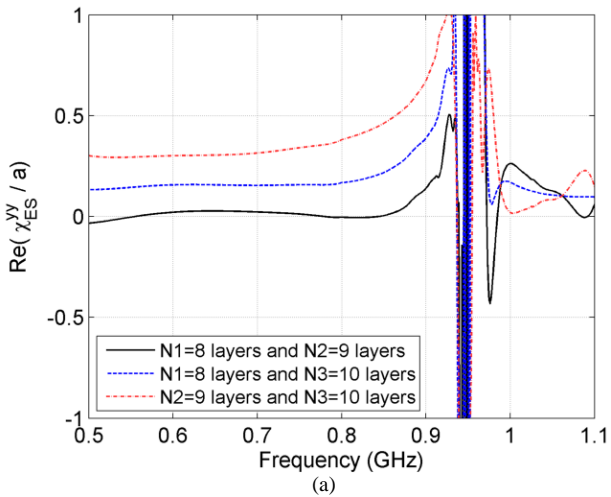

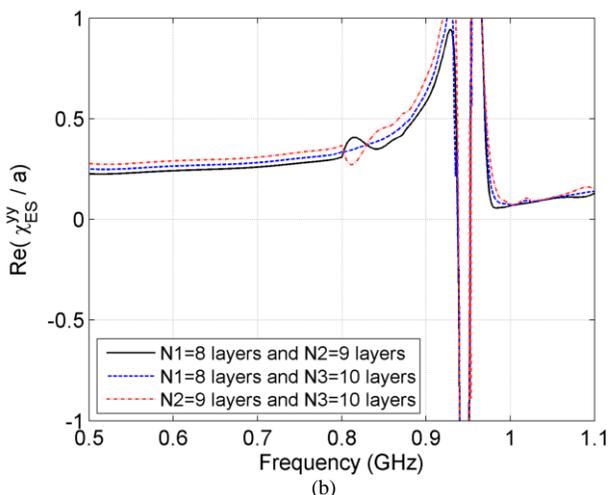

Fig. 9. Normalized electric surface susceptibilities at Interface A or B calculated from our equation for $N1=8$ and $N2=9$ (solid black), $N1=8$ and $N3=10$ (dashed blue), and $N2=9$ and $N3=10$ (dashed red): (a) $\chi_{ES}^{yy}/a$ from the simulation with coarse meshes: max. 1000 meshes for the spheres. (b) $\chi_{ES}^{yy}/a$ from the simulation with fine meshes: max. 2000 meshes for the spheres.

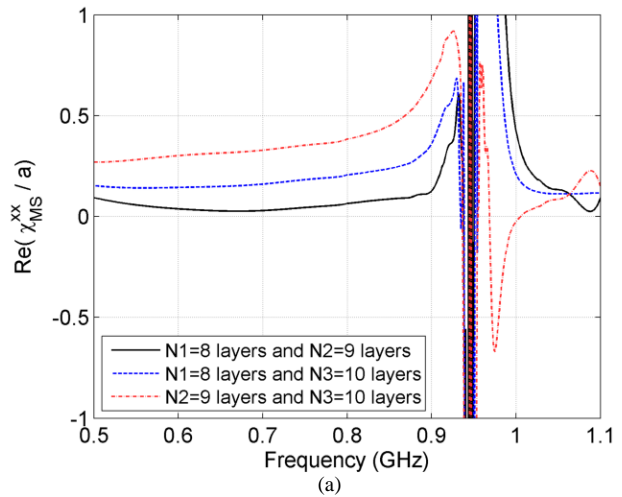

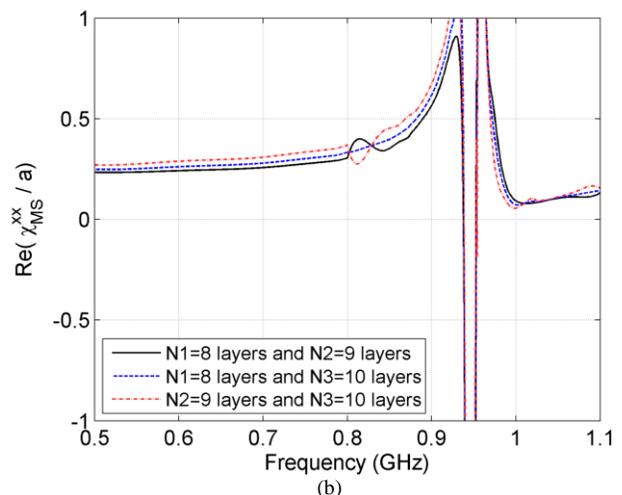

Fig. 10. Normalized magnetic surface susceptibilities at Interface A or B calculated from our equation for $N1=8$ and $N2=9$ (solid black), $N1=8$ and $N3=10$ (dashed blue), and $N2=9$ and $N3=10$ (dashed red): (a) $\chi_{MS}^{xx}/a$ from the simulation with coarse meshes: max. 1000 meshes for the spheres. (b) $\chi_{MS}^{xx}/a$ from the simulation with fine meshes.: max. 2000 meshes for the spheres

calculated from (20) gives 0.3% and 0.6% differences compared to Smith's method at 0.9 GHz (where the refractive index is positive) and 0.95 GHz (where it is negative) for the case when the permittivity and permeability of inclusions are the same: $\varepsilon_r = \mu_r = 50 - j0.05$. Notice that the method of this paper provides the same refractive index regardless of the choice of the reference planes, $L = Na$ or $L = (N-1)a$, while Smith's method requires precise knowledge of the effective slab lengths. For example, if $L = (N-1)a$ is chosen, the real part of the effective refractive index from our approach shows 9% and 13% discrepancies with that from Smith's method at 0.9 and 0.95 GHz. Therefore, one might deduce that $L = Na$ yields the correct choice of reference planes for the property determination techniques used by other researchers up until now; it would certainly seem to be the most natural one. Our effective refractive index is shown to be quite close to that obtained by Scher's method [16]. Note however that the peak values in the real and imaginary parts from that single-layer-based approach are different than those from ours. Scher's method one of modeling rather than retrieval, and is quite independent of the present method.

To validate the independence of the retrieved material properties from the numbers of layers for this metamaterial, we calculated the effective electric and magnetic surface susceptibilities at the interfaces by choosing three different numbers of layers. Our method in this paper assumes that the surface susceptibilities are the same independently of the numbers of layers, as long as there are no field interactions of higher-order (evanescent) modes that arise at the interfaces, and well-converged material properties are ensured. Fig. 9 shows the real parts of the normalized effective electric surface susceptibilities from (7) or (9) for $N1=8$, $N2=9$, and $N3=10$. The surface susceptibilities in Figs. 9(a) and (b) result from S-parameters by coarse (max. 1000 meshes for the spheres) and fine (max. 2000 meshes for the spheres) meshes in the HFSS finite-element simulations and have been normalized by the lattice constant $a$. Note that it has been assumed that the surface susceptibilities are same at the two interfaces, so (7) and (9) give same value. As can be seen in Figs. 9(a) and (b), more accurately simulated S-parameters (finer meshes) provide more converged electric surface susceptibilities for $N1$, $N2$, and $N3$. This indicates that our method may be quite sensitive to errors in measured or simulated S-parameters, rather than that the layer thicknesses used here are not yet large enough to exhibit the convergence of the material properties. Therefore, we consider that good convergence is achieved from 9 layers and 10 layers for the material properties of this metamaterial. It is also shown in the plots that the surface susceptibilities are very noisy around the resonance. Scher *et al.* [26] previously illustrated that in the condition of $\left| n_{eff} \right| \ge 20.2$, there will be "extraordinary" modes that allow more than one mode to propagate simultaneously, which is a consequence of spatial dispersion. This explains our noisy surface susceptibilities around the resonance. Figs. 10(a) and (b) show the real parts of the normalized effective magnetic surface susceptibilities computed from (8) or (10). Similar behavior is observed.

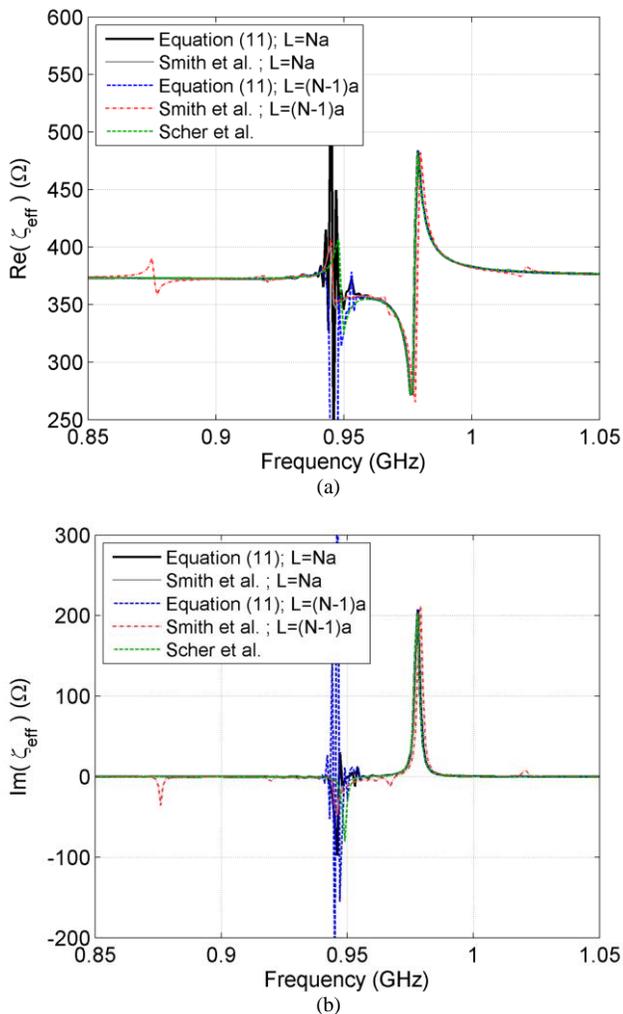

(a)

(b)

Fig. 11. Comparison of the effective wave impedance from (11) (solid black and dashed blue), from the method of Smith *et al.* (solid gray and dashed red) and from the method of Scher *et al.* (dashed green) for $\varepsilon_r = \mu_r = 50 - j0.05$ with $L = Na$ and $L = (N-1)a$: (a) real part of $\varsigma_{eff}$, (b) imaginary part of $\varsigma_{eff}$.

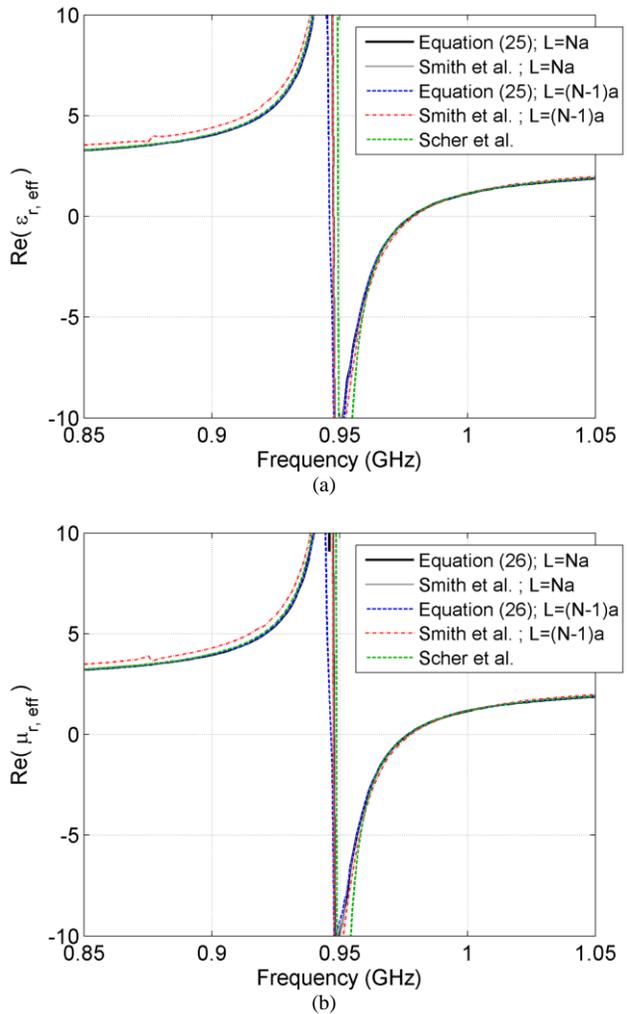

(a)

(b)

Fig. 12. Comparison of the effective relative permittivity and permeability obtained from the present method (solid black and dashed blue), those from the method of Smith *et al.* (solid gray and dashed red) and from the method of Scher *et al.* (dashed green) for $\varepsilon_r = \mu_r = 50 - j0.05$ with $L = Na$ and $L = (N-1)a$: (a) real part of $\varepsilon_{r,eff}$, (b) real part of $\mu_{r,eff}$.

The real and imaginary parts of the effective wave impedance computed from our and Smith's equations are plotted in Figs. 11(a) and (b). Our equation (11) for $\varsigma_{eff}$ was derived from the GSTCs, whereas Smith's formula does not account for boundary effects. The real and imaginary parts of our wave impedance deviate from those of the air between 0.94 GHz and 1.01 GHz. The wave impedance found from our equation has a resonant peak around 0.94-0.953 GHz and 0.977-0.979 GHz.

Figs. 12(a) and (b) show the real parts of the effective relative permittivity and permeability obtained from the same three effective material property extraction methods. The real parts of the effective relative permittivity and permeability from all three exhibit negative values and a resonance around 0.948 GHz. As can be expected, a metamaterial configuration with identical permittivity and permeability of the inclusions results in the same values for the bulk permittivity and permeability of the metamaterial. Our property determination generates the same permittivity and permeability, regardless of whether $L = Na$ or $L = (N-1)a$ is used, whereas Smith's retrieved parameters are significantly different, depending on the choice of the reference planes. As far as $L = Na$ is concerned, our effective property determination taking the boundary effects into consideration results in bulk permittivity and permeability that are approximately 0.4% and 2% different at 0.9 and 0.95 GHz from those from Smith *et al.*

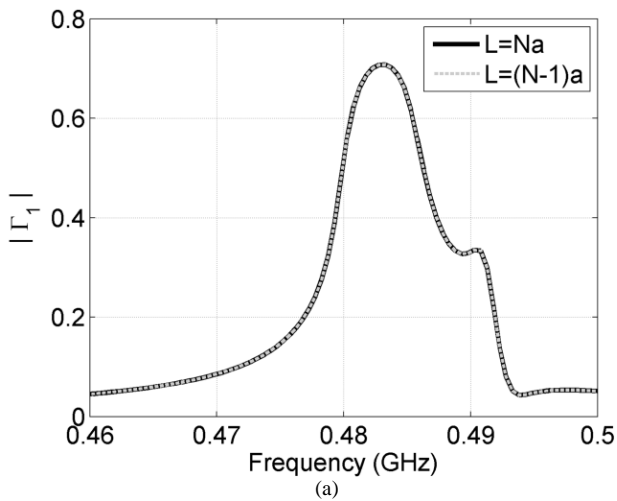

(a)

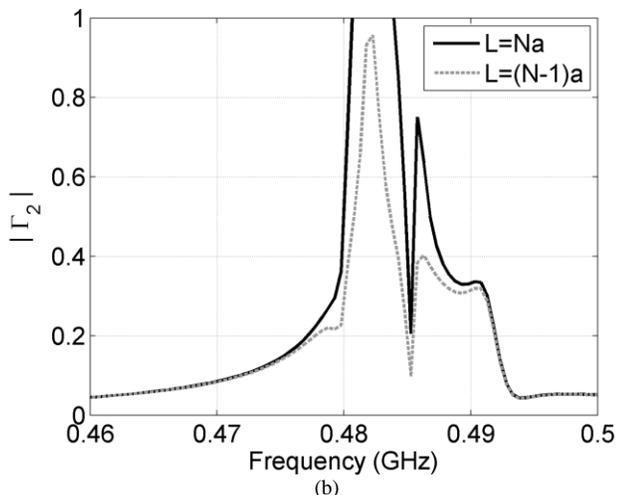

(b)

Fig. 13. Magnitude of the reflection coefficients using our equations for $\varepsilon_r = 130 - j0.26$ and $\mu_r = 75 - j0.15$ with $L = Na$ (solid black) and $L = (N-1)a$ (dashed gray): (a) magnitude of $\Gamma_1$, (b) magnitude of $\Gamma_2$.

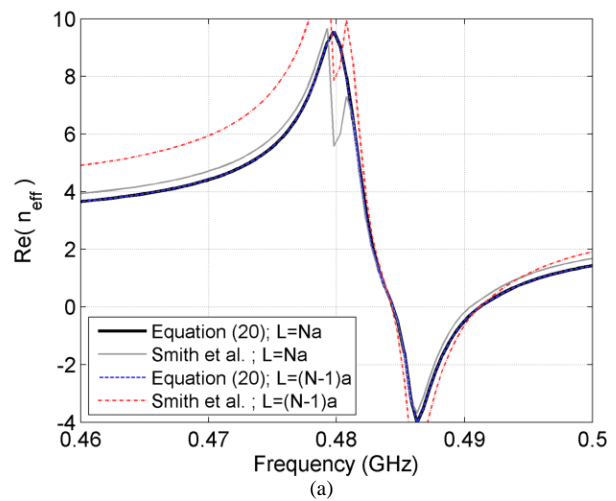

(a)

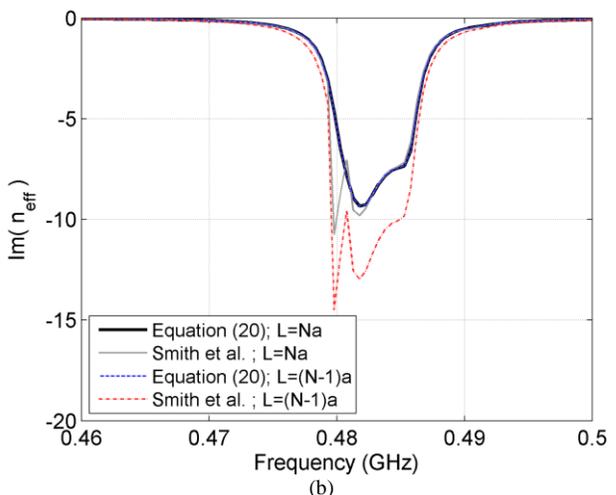

(b)

Fig. 14. Comparison of the effective refractive index by use of (8) (solid black and dashed blue) and the method of Smith *et al.* (solid gray and dashed red) for $\varepsilon_r = 130 - j0.26$ and $\mu_r = 75 - j0.15$ with $L = Na$ and $L = (N-1)a$: (a) real part of $n_{eff}$, (b) imaginary part of $n_{eff}$.

### C. Case2: Simulation Results from a Cubic Array of Magnetodielectric Spheres with $\varepsilon_r \neq \mu_r$

We finally examined the case when the permittivity and permeability of inclusions were very different;

$\varepsilon_r = 130 - j0.26$ and $\mu_r = 75 - j0.15$. For the 3-D full wave simulations, the structure of the metamaterial shown in Fig. 6 was used with two different numbers of layers, $N1 = 4$ and $N2 = 5$. To begin with, we attempted to choose $N1 = 9$ and $N2 = 10$. It however turned out that the simulation generated transmissions so small ($S_{21} < -80$ dB) near the resonance that the retrieval method could not be applied with any accuracy. It would be expected that the boundary effects will be more significant in this case, since the wave impedance of the resultant metamaterial will be different than that of air. Once again, the S-parameter data for a four-layer slab was employed in our implementation of the method of Smith *et al.*

The magnitudes of $\Gamma_1$ and $\Gamma_2$ from (22) and (23) with the choices of the reference planes $L = Na$ and $L = (N-1)a$ are plotted in Figs. 13(a) and (b) as functions of frequency. $|\Gamma_1|$ and $|\Gamma_2|$ are now very different, even more dramatically so than for the metamaterial considered in Section III *B.* $|\Gamma_1|$ is consistent for both $L = Na$ and $L = (N-1)a$. However, $|\Gamma_2|$ shows some dependence on the choice of reference planes near the resonance frequency. This is explained by the fact that as the reference planes are moved, $\Gamma_1$ maintains a constant magnitude while its phase shift is suitably adjusted because the exterior medium is lossless. However, the metamaterial is quite lossy between 4.8 and 4.85 GHz, so $|\Gamma_2|$ shows considerable change with change of reference planes. Note also that $|\Gamma_2|$ exceeds unity for frequencies from 4.82 to 4.84 GHz. This is not incompatible with a passive medium as long as the metamaterial is lossy, and it is speculated that a small stopband may exist at these frequencies wherein the loss is enhanced.

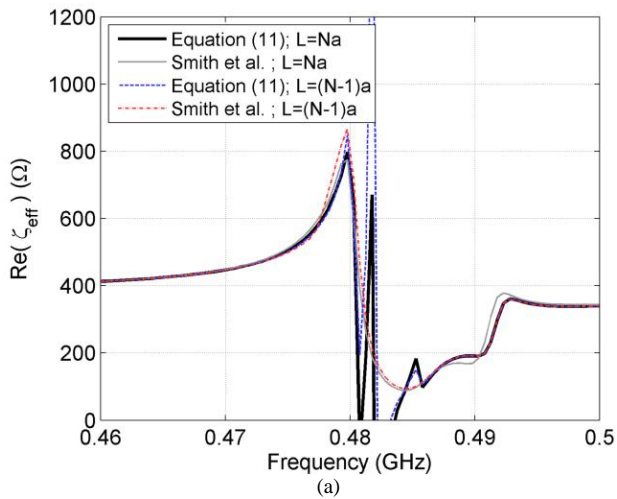

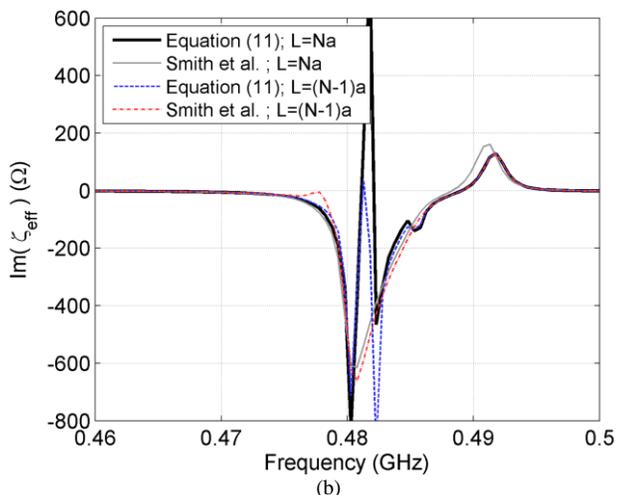

Fig. 15. Comparison of the effective wave impedance using (20) (solid black and dashed blue) and the method of Smith *et al.* (solid gray and dashed red) for $\varepsilon_r = 130 - j0.26$ and $\mu_r = 75 - j0.15$ with $L = Na$ and $L = (N-1)a$: (a) real part of $\varsigma_{eff}$, (b) imaginary part of $\varsigma_{eff}$.

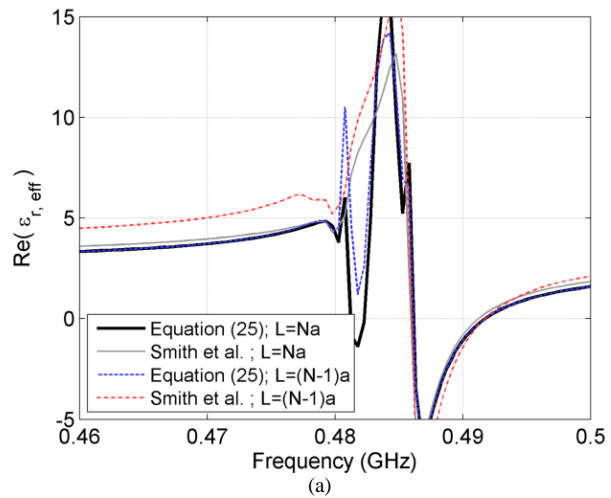

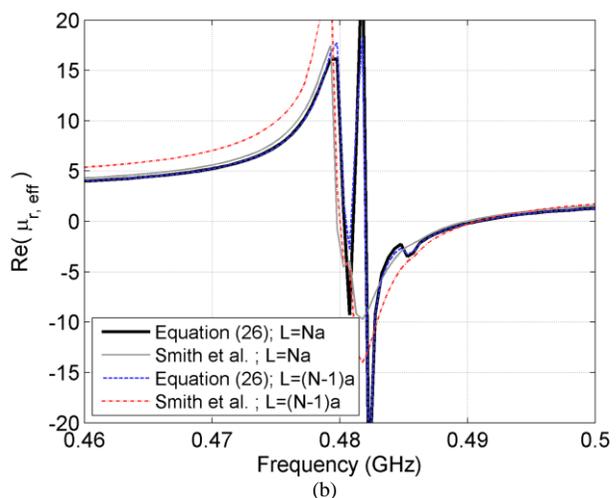

Fig. 16. Comparison of the effective relative permittivity and permeability using our method (solid black and dashed blue) and that of Smith *et al.* (solid gray and dashed red) for $\varepsilon_r = 130 - j0.26$ and $\mu_r = 75 - j0.15$ with $L = Na$ and $L = (N-1)a$: (a) real part of $\varepsilon_{r,eff}$, (b) real part of $\mu_{r,eff}$.

Fig. 14 shows the effective refractive index obtained from (20) and from Smith's method. As expected, our equation gives a consistent effective refractive index even if the reference planes are varied. On the other hand, the result from Smith's equation is dependent on the choice of reference planes, similarly to the previous example. If $L = Na$ is chosen, compared with the results of Smith *et al.*, our approach in this work shows approximately 6% and 23% differences at 4.70 and 4.95 GHz, where the refractive index is positive and negative respectively.

The effective wave impedance is shown in Fig. 15 as a function of frequency. Again our result is the same for $L = Na$ and $L = (N-1)a$, except that our results have different peak values at 0.482 GHz. If $L = Na$ is used for the reference plane, the plots in Figs. 15(a) and (b) indicate that our effective material property determination considering the boundary effects gives a clear discrepancy compared to the values retrieved without it from 0.488 to 0.492 GHz, in which the negative refractive index is observed in Fig. 14(a). One may speculate that the spikes of the real and imaginary parts calculated from our equations are insignificant, since the transmission is very small around those frequencies, according to the plots in Fig. 13.

The real parts of the effective relative permittivity and permeability are shown in Fig. 16. Although our results include the peaks attributed to those of the computed wave impedance, they otherwise show agreement between the results for $L = Na$ and $L = (N-1)a$ . The permittivity becomes negative from 0.486-0.491 GHz and the permeability from 0.481-0.490 GHz. Thus, the metamaterial shows a cutoff property in the frequency range 0.481-0.485 GHz, since only the permeability is negative here. We also observe about a 27% discrepancy for the negative permittivity at 0.49 GHz and 16% discrepancy for the negative permeability at 0.487 GHz between our retrieval method and Smith's method, if the reference plane is taken as $L = Na$ .

Finally, although we do not show the plots for this metamaterial configuration, both our and Smith's approaches show reasonable results for $\mathrm{Im}\left(\varepsilon_{r,\mathit{eff}}\right)$ and $\mathrm{Im}\left(\mu_{r,\mathit{eff}}\right)$ , except for spurious positive values at the single frequency of 0.481 GHz, where $S_{21}$ is very small. Both methods seem to be prone to such anomalies when the transmitted wave is so small as to be essentially just "noise".

## IV. CONCLUSION

We have presented LRL-like expressions for extracting the material parameters of a metamaterial. This is based on the measurement of the S-parameters for two different material sample lengths without the need for any information interior to the samples. The effective refractive index of the metamaterial is found from the S-parameters. The exterior reflection coefficient at the interface is then derived from them. We have also derived an equation for the effective wave impedance with the assumption that GSTCs account for the boundary effects. Initial validation of the method has been done by use of ordinary material (nylon) samples. The results from our equations are in good agreement with those from the NRW algorithm. It was shown that the boundary effects are negligible in this case, and the Fresnel formulas hold at the interfaces.

We should emphasize that boundary effects and spatial dispersion are important phenomena that will occur in a metamaterial. Indeed, boundary effects can be viewed as a consequence of spatial dispersion. The effect of spatial dispersion is included in the values of the effective parameters that we obtain via our method. Since this paper is limited to normally propagating waves, what we extract will be the parameters at a particular frequency and for a normally-directed spatial wavenumber, and cannot be assumed to be the same as those which would apply to waves propagating obliquely. We do not imply that results so obtained will be applicable to any other situation than what was true during the measurement, and in particular to the case of a sample of metamaterial in a waveguide.

Boundary effects on the determination of the effective properties of a metamaterial consisting of a cubic array of magnetodielectric spheres have also been investigated by use of HFSS-simulated S-parameters. Two metamaterial configurations have been tested; one for the case when the permittivity and permeability of the inclusions are identical, and the other for the case of different values of these parameters. In the first case, a resonance at 0.949 GHz is observed, near which the refractive index, permittivity, and permeability are observed to become negative. Comparison with results from Smith's equations shows discrepancies on the order of less than 1%. Our results are also found to be close to those from Scher's method. In addition, the surface susceptibilities have been computed to validate well-converged bulk material properties extracted from the numbers of layers we have chosen. In the second case, a resonance occurs around 4.80-4.86 GHz. A negative index, permittivity, and permeability are also obtained here, and our bulk material property determination taking boundary effects into account yields an effective refractive index and wave impedance with larger differences compared with those using Smith's retrieval method near where negative refractive index occurs (approximately 27% and 16% discrepancies respectively for the negative permittivity and permeability). An important feature of our effective property retrieval method is that the results are independent of the choice of reference planes.

Implicit in our method (as indeed was the case in [7]-[10]) is the assumption that the boundary effect of the metamaterial sample is a local one. In other words, only a single mode of propagation within the metamaterial exists without significant attenuation. Under certain conditions, it has been found that extraordinary modes of propagation with low attenuation may exist in addition to the ordinary one [26], [49]-[50]. In such cases the present method would have to be substantially modified.

It is also noted that from our findings, the transmission coefficients extracted from measured or simulated S-parameters around the resonance frequency of a metamaterial can be very small for the metamaterial models used in this work. Other kinds of extraction equations that are much less sensitive to this (perhaps making use of measured data inside the metamaterial samples) will be necessary in such frequency bands. The related question of sensitivity analysis for our method continues to be investigated.

Finally, we have developed equations for the oblique incidence (TE and TM) cases and thus for metamaterial measurements carried out in a rectangular waveguide. The expressions for an obliquely incident wave (TE/TM) should be used instead of (3)-(6) to take advantage of the GSTCs. The electric and magnetic surface susceptibilities are then found for the oblique incidence case, and finally the wave impedance is found from the surface susceptibilities. Note that the guided wavelength for oblique incidence needs to be used in (15) to find the refractive index. We will report on this in a separate paper, and compare the results from our method and those from another approach [51] based on the Fresnel formulas extended for oblique incidence.

## ACKNOWLEDGMENT

The authors are deeply indebted to Dr. M. D. Janezic of the National Institute of Standards and Technology (NIST) in Boulder, Colorado for the special technical support arrangements. The authors thank Prof. D. S. Filipović at the University of Colorado at Boulder for the use of the HFSS software. The authors thank Prof. C. R. Simovski of Helsinki University of Technology and Saint Petersburg State University of Information Technologies, Mechanics, and Optics for some very fruitful discussions.

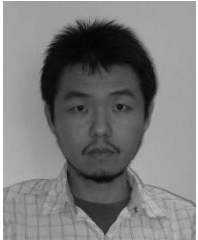

**Sung Kim** (S'08–M'09) received the B.E. degree in electronic engineering from the University of Electro-Communications, Tokyo, Japan, in 1996, the M.S. degree in electrical engineering from California Sate University, Northridge, in 2001, and the Ph.D. degree in electrical engineering from the University of Colorado, Boulder, in 2009.

From 1996 to 1997, he was with Yokowo Co. Ltd., Tokyo, designing LNAs for GPS antennas and millimeter-wave Doppler radars for vehicle sensors. From 2001 to 2003, he worked for Opnext Japan, Inc. (formerly the optical module division of Hitachi, Ltd.), Yokohama, Japan, where he developed 10/40-Gbit/s optical transmitters for optical fiber communications. From 2003 to 2005, he joined Soko Electronics Co. Ltd., Osaka, Japan, as electrical engineer. He is currently with the Electromagnetics Division, the National Institute of Standards and Technology (NIST), Boulder, as a postdoctoral researcher.

His research interests include the measurements and designs of microwave and optical components.

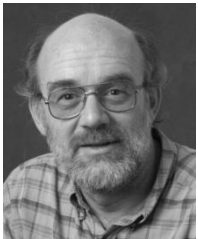

**Edward F. Kuester** (S'73–M'76–SM'95–F'98) received the B.S. degree from Michigan State University, East Lansing, in 1971 and the M.S. and Ph.D. degrees from the University of Colorado, Boulder, in 1974 and 1976, respectively, all in electrical engineering.

Since 1976, he has been with the Department of Electrical and Computer Engineering, University of Colorado, where he is currently a Professor. In 1979, he was a Summer Faculty Fellow at the Jet Propulsion Laboratory, Pasadena, CA. During 1981–1982, he was a Visiting Professor at the Technische Hogeschool, Delft, The Netherlands. During 1992–1993, he was Professeur Invité at the École Polytechnique Fédérale de Lausanne, Switzerland. He was a Visiting Scientist at the National Institute of Standards and Technology (NIST), Boulder, in 2002, 2004, and 2006. His current research interests include the modeling of electromagnetic phenomena of guiding and radiating structures, applied mathematics, and applied physics. He is the coauthor of one book, the author of chapters in two others, and translator of two Russian books. He is coholder of two U.S. patents and author or coauthor of more than 60 papers in refereed technical journals and numerous conference presentations.

Dr. Kuester is a Fellow of the IEEE and (AP), IEEE Microwave Theory and Techniques (MTT), and IEEE Electromagnetic Compatibility (EMC) Societies, a member of the Society for Industrial and Applied Mathematics, and a member of Commissions B and D of the International Union of Radio Science.

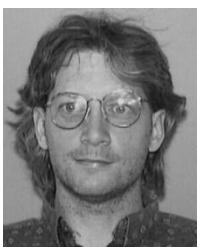

**Christopher L. Holloway** (S'86–M'92–SM'04 – F'10) was born in Chattanooga, TN, on March 26, 1962. He received the B.S. degree from the University of Tennessee, Chattanooga, in 1986 and the M.S. and Ph.D. degrees in electrical engineering from the University of Colorado, Boulder, in 1988 and 1992, respectively.

During 1992, he was a Research Scientist with Electro Magnetic Applications, Inc., Lakewood, CO, where he was engaged in theoretical analysis and finite-difference time-domain modeling of various electromagnetic problems. From 1992 to 1994, he was with the National Center for Atmospheric Research (NCAR), Boulder, where he was engaged in wave-propagation modeling, signal-processing studies, and radar-systems design. From 1994 to 2000, he was with the Institute for Telecommunication Sciences (ITS), U.S. Department of Commerce, Boulder. During this period, he was working on wave propagation studies. Since 2000, he has been with the National Institute of Standards and Technology (NIST), Boulder, working on electromagnetic theory. He is also a Graduate Faculty member at the University of Colorado. His current research interests include electromagnetic field theory, wave propagation, guided wave structures, remote sensing, numerical methods, and electromagnetic compatibility/electromagnetic interference (EMC/EMI) issues. He is an Associate Editor of the IEEE TRANSACTIONS ON ELECTROMAGNETIC COMPATIBILITY.

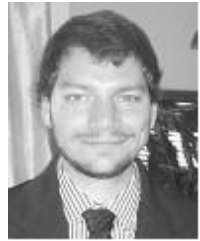

**Aaron D. Scher** was born in Seattle, WA on May 27, 1981. He received the B.S. and M.S. degrees from Texas A&M University, College Station, in 2003 and 2005, respectively, and the Ph.D. degree from the University of Colorado, Boulder, in 2008, all in electrical engineering. He is currently working as a postdoctoral researcher at the University of Colorado. His research interests include the characterization and analysis of artificial composites and applied electromagnetics.

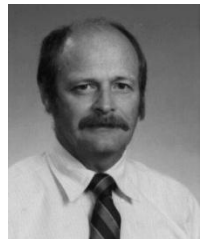

**James Baker-Jarvis** (M'89–SM'90–F'10) was born in Minneapolis, MN, in 1950, and received the B.S. degree in mathematics in 1975. He received the Masters degree in physics in 1980 from the University of Minnesota and the Ph.D. degree in theoretical physics from the University of Wyoming in 1984.

He worked as an AWU Postdoctoral Fellow after graduation for one year on theoretical and experimental aspects of intense electromagnetic fields in lossy materials and dielectric measurements. He then spent two years as an Assistant Professor with the Physics Department, University of Wyoming, working on electromagnetic heating processes and taught classes. Through 1988, he was an Assistant Professor of Physics with North Dakota State University (NDSU). At NDSU, he taught courses in the areas of electronic properties of materials and performed research on an innovative technique to solve differential equations using a maximum-entropy technique. He joined the National Institute of Standards and Technology (NIST), Boulder, in January 1989 where he has worked in the areas of theory of microscopic relaxation, electronic materials, dielectric and magnetic spectroscopy, and nondestructive evaluation. He is Project Leader of the Electromagnetic Properties of Materials Project at NIST. He is the author of numerous publications. His current interests are in dielectric measurement metrology, theoretical microscopic electromagnetism, and quantum mechanics.

Dr. Baker-Jarvis is a NIST Bronze Medal recipient.